\documentclass[
prl,aps,
reprint,
a4paper,
amsmath,amssymb,amsbsy,
floatfix]{revtex4-1}

\usepackage{etoolbox}
	\newtoggle{arXiv}
	\toggletrue{arXiv}

\usepackage{latexsym}
\usepackage{graphics}
\usepackage{amssymb}
\usepackage{amsmath}
\usepackage{multirow}
\usepackage{textcomp}
\usepackage{siunitx}
\usepackage{bbold}
\usepackage{relsize}
\usepackage{bbm}
\usepackage{float}

\newcommand{\be}{\begin{eqnarray}}
\newcommand{\ee}{\end{eqnarray}}
\newcommand{\ket}[1]{\ensuremath{\left| {#1} \right>}}
\newcommand{\bra}[1]{\ensuremath{\left< {#1} \right|}}
\newcommand{\ketS}[1]{\ensuremath{| {#1} \rangle}}
\newcommand{\braS}[1]{\ensuremath{\langle {#1}|}}

\newcommand{\ketbra}[2]{\ensuremath{| {#1} \rangle \langle {#2} |}}

\newcommand{\create}{\ensuremath{{\,\hat{a}^{\dagger}}}}
\newcommand{\destroy}{\ensuremath{\,\hat{a}}}

\newcommand{\Dis}[1]{\hat{\mathcal{D}}(#1)}

\usepackage{tikz}
\usetikzlibrary{arrows,calc,matrix,fit,positioning}
\usetikzlibrary{decorations.pathreplacing}
\usetikzlibrary{decorations.pathmorphing}
\usetikzlibrary{decorations.markings}
\usepackage{quantumcircuits}

\tikzset{
	circuit/.style = {
		row sep=1ex,
		column sep=2em,
		nodes = {anchor=center},
	},
	gate/.style = {minimum width=7mm,minimum height=7mm,draw},
	Bgate/.style = {minimum width=10mm,minimum height=7mm},
	ctrl/.style = {circle,fill=black,inner sep=0.7mm},
	NOT/.style = {pluscircle,draw},
	meas/.style = {minimum width=7mm,minimum height=7mm,measurement,draw},
	state/.style = {},
	quantum bit/.style = {},
	classical bit/.style = {double=white,draw=black,double distance=0.6mm,line width=0.3mm},
}

\def\gate#1{|[gate]|#1}

\def\C{|[ctrl]|}

\def\meas{|[meas]|}

\begin{document}

\title{Sequential modular position and momentum measurements of a trapped ion mechanical oscillator}

\author{C. Fl{\"u}hmann} \email{christaf@phys.ethz.ch}
\author{V. Negnevitsky}
\author{M. Marinelli}
\author{J.~P.~Home}\email{jhome@phys.ethz.ch}
\affiliation{Institute for Quantum Electronics, ETH Z\"urich, Otto-Stern-Weg 1, 8093 Z\"urich, Switzerland}


\begin{abstract}
The non-commutativity of position and momentum observables is a hallmark feature of quantum physics.
However this incompatibility does not extend to observables which are periodic in these base variables. Such modular-variable observables have been suggested as tools for fault-tolerant quantum computing and enhanced quantum sensing.
Here we implement sequential measurements of modular variables in the oscillatory motion of a single trapped ion, using state-dependent displacements and a heralded non-destructive readout. We investigate the commutative nature of modular variable observables by demonstrating no-signaling-in-time between successive measurements, using a variety of input states. In the presence of quantum interference, which we enhance using squeezed input states, measurements of different periodicity show signaling-in-time. The sequential measurements allow us to extract two-time correlators for modular variables, which we use to violate a Leggett-Garg inequality. The experiments involve control and coherence of multi-component superpositions of up to 8 coherent, squeezed or Fock state wave-packets. Signaling-in-time as well as Leggett-Garg inequalities serve as efficient quantum witnesses which we probe here with a mechanical oscillator, a system which has a natural crossover from the quantum to the classical regime.
\end{abstract}


\maketitle
One of the fundamental notions of quantum mechanics is that position and momentum operators do not commute. This restricts the possible states a particle can be prepared in to fulfill the Heisenberg uncertainty principle: $\Delta \hat{x} \Delta \hat{p} \geq \frac{1}{2} \left|\left\langle \left[\hat{x}, \hat{p} \right]  \right\rangle \right|$ with $\left[\hat{x}, \hat{p} \right] = i \hbar$, and limits the ability to perform simultaneous position and momentum measurements \cite{03Ozawa,04Ozawa,04Hall,17Terhal}. However this is different for measurements of position and momentum modulo a characteristic length/momentum scale (i.e. $\hat{X} \text{ mod } l_x$, $\hat{P} \text{ mod } l_p$), which can commute.
Such variables were first discussed in the context of the seminal Aharonov-Bohm effect \cite{Aharonov1969} and provide new perspectives in the study of fundamental aspects of quantum mechanics. For example, they exhibit non-local Heisenberg equations of motion \cite{10Popescu}. Modular variables have been proposed for testing macro-realism via Leggett-Garg Inequalities (LGI) \cite{15Asadian} as well as contextuality with continuous-variable systems \cite{15Asadian2}.
The commutation of modular position and momentum allows their use as stabilizers for fault-tolerant continuous variable computation, as proposed by Gottesman, Kitaev and Preskill (GKP) \cite{01Perskill}. Additionally, sequences of these measurements have been proposed to prepare approximate GKP code states \cite{16Weigand,02Travaglione}.
In contrast, for incompatible modular position and momentum measurement settings we expect the first measurement to influence the statistics of the subsequent measurement, which has previously been defined as Signaling-In-Time (SIT) \cite{13Kofler,12li}.  Observation of SIT and LGI violations provide means to exclude macro-realistic theories and often serve as quantum witnesses \cite{13Kofler,12li,14Emary,85Leggett}.

In this Letter we implement and analyze sequences of modular position and momentum measurements of a quantum harmonic oscillator realized in the axial motional oscillation of a single trapped calcium 40 atomic ion. The observables are measured by coupling the oscillator to the ion's internal qubit states using state-dependent forces, and subsequently reading out the qubit using resonance fluorescence \cite{98Wineland2}. We analyze SIT between the measurements and violate a LGI. Using both methods we confirm the quantum nature of the motional states using a small number of measurements. In addition we test the commutation of modular measurements by observing Non-Signaling-In-Time (NSIT) on a variety of input states.

In a first set of measurements, we perform ``symmetric'' modular measurements using a bi-chromatic laser field resonant with both the red and blue sideband of the quadrupole transition between the $\ket{\downarrow} \equiv \ket{S_{1/2}, m_j=1/2}$ and $\ket{\uparrow} \equiv \ket{D_{5/2}, m_j=3/2}$ internal states \cite{04Haljan}. This realizes a Hamiltonian $\hat{H}_{\text{SDF}}=\eta \hbar \Omega \hat{ \sigma}_x(\hat a e^{i\Delta\phi/2}+\hat{ a}^{\dagger}e^{-i\Delta\phi/2})/2$, where $\hat{\sigma}_x \equiv \ketbra{\uparrow}{\downarrow} + \ketbra{\downarrow}{\uparrow}$, $\eta \simeq 0.05$ is the Lamb-Dicke parameter \cite{98Wineland2}, $\hat{a}$ is the harmonic oscillator destruction operator and $\Omega$, $\Delta\phi$ are related to the intensity and relative phases of the sideband laser fields. The corresponding time evolution operator is $\hat{\mathcal{D}}(\alpha(t)\hat{ \sigma}_x/2 )$ where $\alpha(t)=i e^{i\Delta \phi/2}\eta \Omega t$ and $\hat{\mathcal{D}}$ is the phase-space displacement operator \cite{05Schleich}. For an initial state $\ket{\downarrow}\otimes \ket{\psi_{\rm in}}$, this results in a qubit-motion entangled state $\ket{\uparrow}\otimes \hat{E}_{-}\ket{\psi_{\rm in}} +  \ket{\downarrow}\otimes \hat{E}_{+}\ket{\psi_{\rm in}}$ with $\hat{E}_{\pm}(\alpha)=(\hat{\mathcal{D}} (-\alpha/2) \pm \hat{\mathcal{D}} (\alpha/2))/2$. The subsequent measurement of the internal state gives the results $\ket{\downarrow}$, $\ket{\uparrow}$ with probability $P(\downarrow/\uparrow)\equiv P(\pm 1) = \text{Tr}(\hat{E}_\pm^{\dagger}\hat{E}_\pm\ketbra{\psi_{\text{in}}}{\psi_{\text{in}}})$, with the corresponding modular measurement operator \cite{15Asadian}
\be
\hat{Q}(\alpha)=\cos(2 \text{Im}(\alpha) \hat{X} - 2\text{Re}(\alpha) \hat{P})
\ee
defined using $\hat{X}=\sqrt{\frac{m\omega}{2\hbar}}\hat{x}$ and $\hat{P}=\sqrt{\frac{1}{2 m\omega \hbar}}\hat{p}$ as dimensionless position and momentum operators, with $\omega \approx 2 \pi \times 1.85 \rm{\:MHz}$ and $m\approx 40 \rm{\:amu}$ denoting the harmonic oscillator frequency and mass. For this definition $[\hat{X},\hat{P}]=i/2$.  By choosing $\alpha$ to be real (imaginary) we perform a modular momentum (position) measurement with modularity dependent on $\alpha$.

In an ideal scenario, the measurement would project into the state $\ket{\psi_\pm} \propto \hat{E}_{\pm}\ket{\psi_{\rm in}}$ conditional on the measurement result. In practice, we measure the qubit using state-dependent resonance fluorescence, which for the detection of $\ket{\uparrow}$ (no photons scattered) closely realizes the ideal scenario. However, measuring the $\ket{\downarrow}$ state involves scattering of around 1000 photons which randomizes the oscillator state. We thus perform the measurement in a heralded fashion, and only analyze the subsequent state (or continue to further measurements) if the detection is dark. This decision is made in real-time using an FPGA to save data acquisition time. In half of our experiments we invert the qubit prior to the fluorescence detection, allowing projection into $\ket{\psi_+} \propto \hat{E}_{+}\ket{\psi_{\text{in}}}$ as a dark measurement result.

In quantum mechanics, the measurement of one quantity often influences a subsequent measurement of a different quantity. We consider two symmetric modular variable measurements $A, B$ with measurement settings controlled through the respective displacements $\alpha_A$ and $\alpha_B$. The measurement outcomes are $a, b \in \{+1, -1\}$. Measurement $B$ is either performed alone or subsequent to a measurement of $A$, resulting in probabilities $P_B(b)$ or $P_{B(A)}(b) \equiv \sum_a P_{BA}(b, a)$ respectively, where we have defined the joint probability $P_{BA}(b,a) \equiv P_{A}(a)P_{B|A}(b|a)$.
For some settings, the statistics of measurement $B$ change if measurement $A$ is performed before it (SIT), while for other settings the statistics of measurement $B$ do not depend on the presence of measurement $A$ (NSIT). Since the measurement is binary, we can quantify SIT of $A$ to $B$ using $S = P_{B}(b = +1)-P_{B(A)}(b = +1)$, which for our experiments results in
\be
S = \frac{1}{2}\left(1-\cos(\Phi)\right) |m_{\alpha_B}| \cos(\arg(m_{\alpha_B}))
\ee
where $m_{\alpha}\equiv \bra{\psi_{\rm in}} \Dis{\alpha} \ket{\psi_{\rm in}} $ and $\Phi = \text{Im}(\alpha_A^* \alpha_B)$ is the geometric phase which arises from the non-commutation of the displacement operators $\Dis{\alpha_A} \Dis{\alpha_B} = e^{i\Phi}\Dis{\alpha_A + \alpha_B}$\cite{05Schleich}. From this expression we see that SIT will not occur for any state if either the geometric phase $\Phi= 2 \pi k$, $k \in \mathbb{Z}$ or the wave packet overlap $|m_{\alpha_B}|=0$. We analyze these dependencies in two experiments. 

\begin{figure}[t]
	\resizebox{0.47\textwidth}{!}{\includegraphics{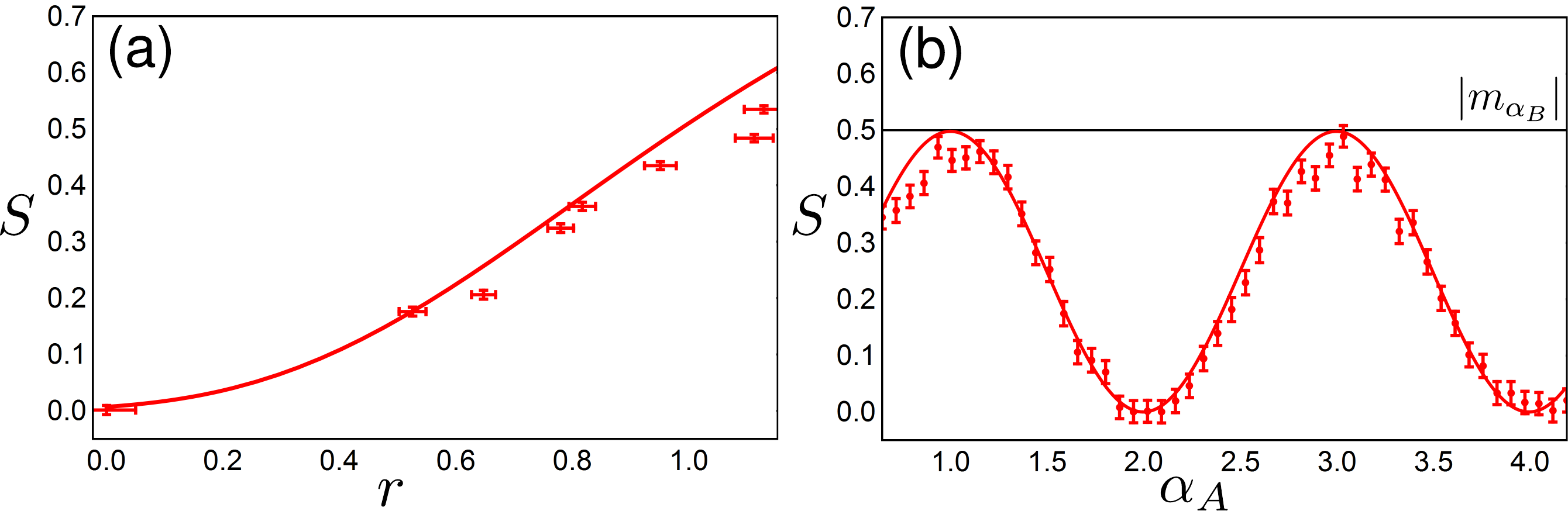}}
	\caption{Dependence of SIT of measurement $A$ to measurement $B$ on (a) interference and (b) geometrical phases. Solid lines show the expectations for an ideal experimental system and the error bars of $S$ are propagated from the shot noise standard errors of the mean (SEM) of the directly measured probabilities $P_B(b), P_A(a)$ and $P_{B|A}(b|a)$. (a): The SIT measurement settings $\alpha_B=3.1 i$, $\alpha_A = 3.02\approx3\pi / |\alpha_B|$ are applied to squeezed input states $\hat{S}(r)\ket{0}$ where the squeezed axis was aligned with position. 
(b): The geometric phase is varied by sweeping the displacement amplitude $\alpha_A$ of measurement $A$, for a ``Schr{\"o}dinger's cat'' input superposition $(\Dis{-\alpha_B/2} +\Dis{\alpha_B/2)}\ket{0}$ with $\alpha_B=i\pi$. 
}
	\label{fig:SIT}
\end{figure}

In the first we examine the effect of wave packet overlap using squeezed vacuum states $\ket{\psi_{\rm in}} = \hat{S}(r)\ket{0}$ with $\hat{S}(r) = e^{r (\destroy^2 - \create^2)/2}$ and where the phase is chosen such that the squeezing parameter $r$ is real and positive. These states can be readily prepared using reservoir engineering \cite{16Kienzler}. We choose the measurement displacement $\alpha_B=3.1 i$, which is aligned with the anti-squeezed axis of the input state. By varying $r$, we can control the wave packet interference, which in this case scales as $m_{\alpha_B} = e^{-|\alpha_B|^2 e^{-2r}/2}$ \cite{15Lo}. We choose $\alpha_A = 3.02 \approx 3\pi/|\alpha_B|$ to ensure that for a given overlap maximal SIT is observed. Experimental results are shown in figure \ref{fig:SIT} (a), exhibiting agreement with the ideal theoretical expectation. Deviations between the two for large $r$ are primarily due to imperfect squeezed state preparation.

The input state $\ket{\psi_{\text{in}}} = (\hat{\mathcal{D}}(-\alpha_B/2) + \hat{\mathcal{D}}(\alpha_B/2))\ket{0}$ exhibits a constant non-zero level of interference $|m_{\alpha_B}|\approx 1/2$. Thus we use this state to illustrate the dependence of SIT on the geometric phase. This is done by setting $\alpha_B=3.1 i$and varying $\alpha_A$ which is taken to be real. Data is shown in figure \ref{fig:SIT} (b) showing oscillations of $S$ with amplitude $|m_{\alpha_B}|$. These oscillations illustrate the periodic effect of the geometric phase. NSIT is seen for this measurement when $\alpha_A = 2\pi k/|\alpha_B|\approx 2 k$.

If NSIT is observed for all possible input states, then it follows that the underlying observables commute. The converse is not true for the non-projective measurements considered here (see Supplemental Information (SI)).
The commutation of observables is hard to verify in practice given the infinite nature of the harmonic oscillator Hilbert space. As a reduced investigation, we examine this property using 150 input states of the form $\ket{\psi_{\text{in}}} = (\hat{\mathcal{D}}(-|\alpha_B|e^{i\phi_I}/2) + \hat{\mathcal{D}}(|\alpha_B| e^{i \phi_I}/2))\ket{\phi}$, where $\ket{\phi}$ is chosen to be one of $(i)$ the ground state $\ket{0}$, $(ii)$ a squeezed state $\hat{S}(-0.82)\ket{0}$ or $(iii)$ the first excited state $\ket{1}$, and for each $\ket{\phi}$ 50 values of $\phi_I$ evenly spaced between zero and $2\pi$ are used. To investigate the commutation of modular position and momentum for large displacements we choose the NSIT geometric phase with $k = 2$ ($\Phi \approx 4 \pi$), which we implement using the measurement settings: $\alpha_B = i\pi$, $\alpha_A = 4.09$.

\begin{figure}[H]
	\resizebox{0.47\textwidth}{!}{\includegraphics{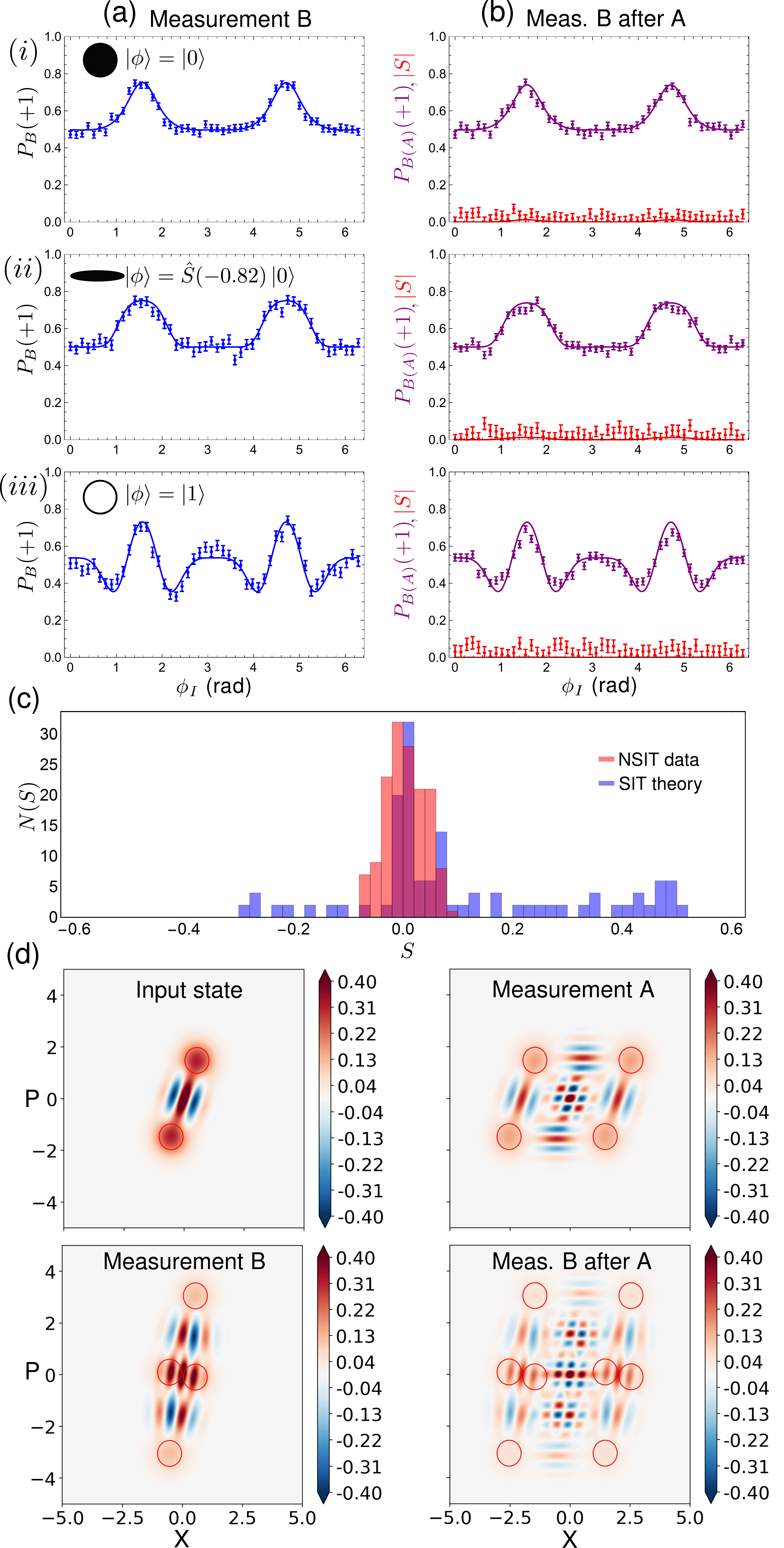}}
	\caption{NSIT of modular position and momentum measurements for a variety of input states. Row $(i)-(iii)$ show NSIT of $A$ to $B$ for the measurement settings $\alpha_B=i\pi$, $\alpha_A = 4.09 \approx 4\pi / |\alpha_B|$, and input superposition states $(\hat{\mathcal{D}}(-|\alpha_B|e^{i\phi_I}/2) + \hat{\mathcal{D}}(|\alpha_B| e^{i \phi_I}/2))\ket{\phi}$ with the phase $\phi_I$ varied. $\ket{\phi}$ is chosen to be $(i)$: the ground state $\ket{0}$, $(ii)$: a squeezed vacuum state $\hat{S}(-0.82)\ket{0}$ or $(iii)$: the first excited state $\ket{1}$. We observe qualitative agreement between column (a), showing measurement of B alone and column (b) showing $B$ measured after $A$. Solid lines show the expectations of an ideal experiment. Errors are given as SEM and propagation of SEM. The 150 measured $S$ values are quantified in red in histogram (c) and compared to a theoretical histogram for the SIT settings  $\alpha_A = 3$ and $\alpha_B = i\pi$ using the same set of input states. (d) Theoretical Wigner function plots of the input state ($\ket{\phi}$=$\ket{0}$ and $\phi_I=1.22 \text{ rad}$) as well as its post-measurement states with result +1 during the experimental sequence. The red circles show the locations of the multiple displaced coherent states, their radius denotes the r.m.s wave-packet size.}
	\label{fig:NSIT}
\end{figure}

Data and a histogram of all measured values of $S$ are shown in figure \ref{fig:NSIT} (a)-(c). For comparison in (c), we also plot theoretical calculations for $\alpha_B = i\pi$, $\alpha_A = 3$, resulting in $\Phi = 3\pi$ which corresponds to maximal SIT but with the same $m_{\alpha_B}$ as used in the experiment. The maximal $|S|$ value measured is $0.087 \pm 0.003$ while the maximum calculated is 0.5. Additionally the standard deviation of the SIT theory histogram is 5.5 times larger than that of the experimentally measured distribution. Theoretical Wigner function plots for one input state example ($\ket{\phi}$=$\ket{0}$ and $\phi_I=1.22 \text{ rad}$) through the experimental sequence are shown in figure \ref{fig:NSIT} (d). The created states are superpositions of up to 8 displaced $\ket{\phi}$ states with separations of up to $\Delta \alpha \approx 8.3$. These measurements illustrate the high level of control for the implemented sequential modular measurements. The ability to tune them from SIT to NSIT demonstrates the quantum nature of the created states and additionally confirms the possibility of modular position and momentum measurements to commute.

An additional means by which successive measurements can be related to one another is through the correlation function of the measurement results, which is defined by $C_{AB} = \sum_{a,b} ab P_{BA}(b,a)$. For the measurements described above, the correlation function between the two measurements is $C_{AB} =  \left(|m_{\alpha_A - \alpha_B}|\cos(\varphi_-) + |m_{\alpha_A + \alpha_B}|\cos(\varphi_+)  \right)/2$ with $\varphi_{\pm}=\arg(m_{\alpha_A \pm \alpha_B})$. This is independent of the geometric phase $\Phi$. 

The multiplication of the Kraus operators by an arbitrary unitary $\hat{U}$ leads to new Kraus operators $\hat{F}_{\pm}=\hat{U}\hat{E}_{\pm}$ but produces the same modular measurement operator $\hat{Q}$. We explore in the following the ``asymmetric'' modular measurement implementation $\hat{F}_{\pm}(\phi,\alpha)=\frac{1}{2}(\mathbb{1} \pm e^{i \phi}\Dis{\alpha})$ which corresponds to $\hat{U}=\Dis{\alpha/2}$. Furthermore we add flexibility to our measurement by controlling the relative phase $\phi$ between un-displaced and displaced components. The generalized observable is then $\hat{Q}(\phi,\alpha)=\cos(\phi+2 \text{Im}(\alpha) \hat{X} - 2\text{Re}(\alpha) \hat{P})$. This is experimentally achieved using a third energy level in the ion (see SI). For this asymmetric implementation we find $\widetilde{S} = \sin(\Phi) |m_{\alpha_B}| \sin(\Phi + \phi_B + \arg{(m_{\alpha_B})})$
and the correlation function is
\be
&&\widetilde{C}_{AB} = \frac{1}{2} \left(|m_{\alpha_A - \alpha_B}|\cos(\widetilde{\varphi}_-) + |m_{\alpha_A + \alpha_B}|\cos(\widetilde{\varphi}_+)  \right)\notag\\
&&\widetilde{\varphi}_{\pm}=\phi_A \pm \phi_B \pm \Phi + \arg({m_{\alpha_A \pm \alpha_B}}).
\ee
This implementation reintroduces the geometric phase to the correlator.

\begin{figure}[t]
	\resizebox{0.47\textwidth}{!}{\includegraphics{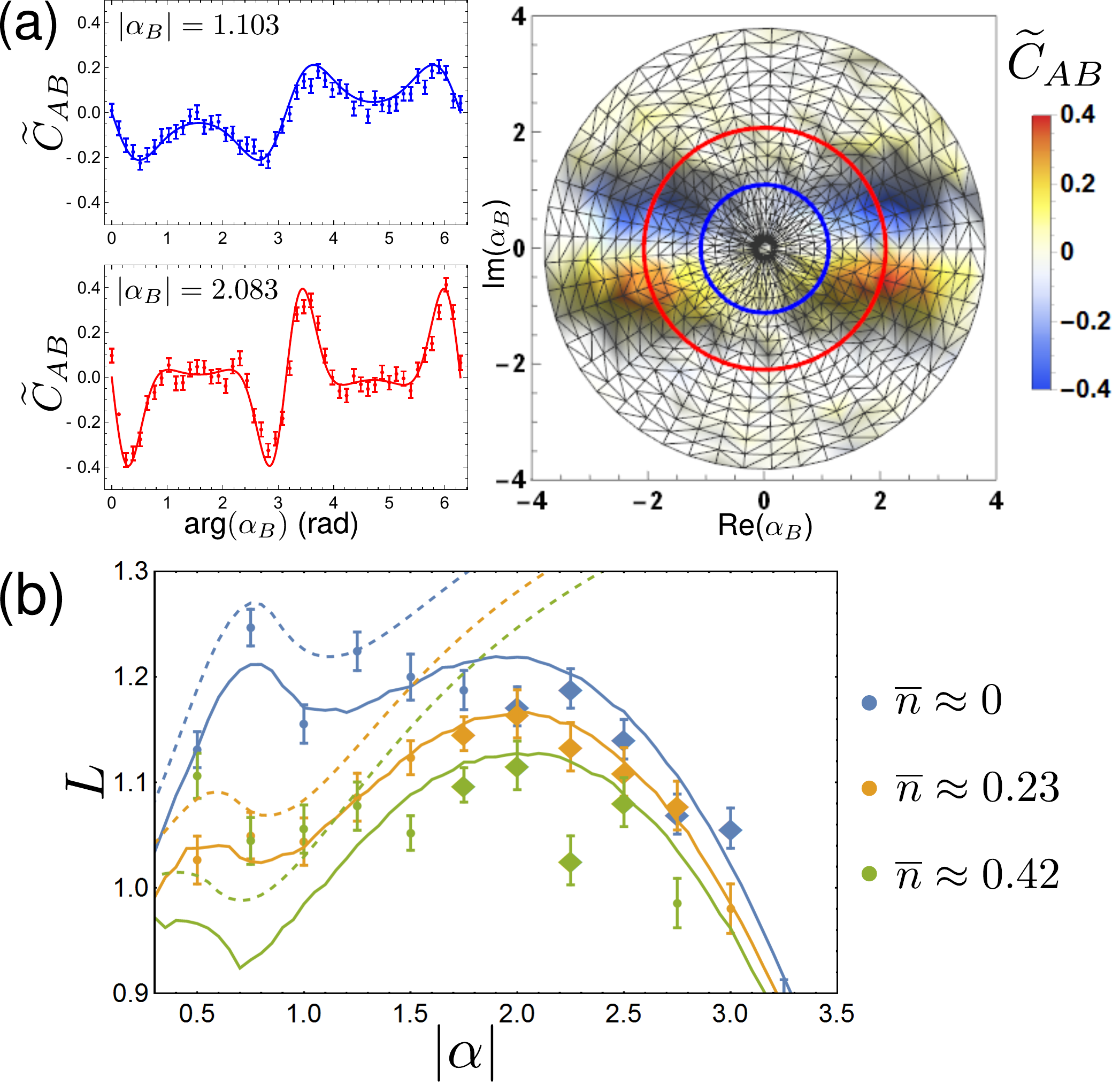}}
	\caption{Two time correlators and violation of a Leggett-Garg inequality. (a) Two-time correlation measurement in the asymmetric implementation performed on an initial ground state cooled ion as a function of the second displacement $\alpha_B$. The fixed experimental settings were: $\alpha_A=2.1$, $\phi_A=0$, $\phi_B=\pi/2$. The full data set is shown in the density plot to the right. Two cuts through this data set with fixed $|\alpha_B|$ are indicated in this density plot and explicitly plotted to the left, where solid lines show the expectations for an ideal experiment. (b) Violation of a Leggett-Garg inequality for increased modular measurement displacements $|\alpha|$ and 3 different initial temperatures. Solid lines show the expected violation including simulated qubit, and motional dephasing as well as phase calibration errors (see SI), the dashed lines instead are the exceptions for an ideal experiment. Violations are observed over a wide range of $\alpha$ for all investigated temperatures. With a ground state cooled ion violations are observed up to $\alpha=3$. The points highlighted with a diamond violate the LGI when being penalized by the inbuilt theoretical amount of SIT. The discrepancy between the data and the simulation at $\overline{n}=0.42$ is due to additional experimental fluctuations in the preparation of this higher thermal occupation. All error bars are propagated from the SEM errors due to quantum projection noise.}
	\label{fig:Correlation}
\end{figure}

A measurement of the correlation function using the asymmetric implementation as a function of $\alpha_B$ is shown in figure \ref{fig:Correlation} (a). The experimental parameters were $\alpha_A=2.1$, $\phi_A=0$, $\phi_B=\pi / 2$ and the input state was the ground state $\ket{\psi_{\rm in}} = \ket{0}$. 
The correlation function then reads $\widetilde{C}_{AB} = -(e^{-|2.1 - \alpha_B|^2/2}+e^{-|2.1 + \alpha_B|^2/2})\sin(\Phi)/2$ with the geometric phase $\Phi=2.1\text{Im}(\alpha_B)$. The pre-factor is non-zero for $\alpha_B \approx \pm 2.1$, in this case wave-packets overlap in the post-measurement state of $B$ leading to interference effects during the measurement. The sign change of the correlator across the real axis is solely due to the geometric phase $\Phi$. The extreme values of $\widetilde{C}_{AB}$ are reached as a compromise between the wave-packet overlap and the geometric phase.

Correlation functions lie at the heart of many tests of the quantum nature of physical systems \cite{67Kochen,15Hensen,85Leggett}. For systems measured at sequential times, the best known is the Leggett-Garg inequality, for which one form is given by
\be
L = C_{A B} + C_{B C} - C_{A C} \leq 1
\ee
where a time sequence of three measurements $A,B,C$ is considered. The bound is derived under two assumptions, which are that the measurement results are of macroscopic quantities which are pre-determined in advance of the experiment, and that they are unchanged by the act of measurement \cite{85Leggett}. Therefore to exclude macroscopic realism Non-Invasive Measurements (NIM) need to be used in the experiment, which is hard to ensure in practice.

A protocol for testing LGI using modular variable measurements has been proposed previously \cite{15Asadian}. There, it was shown that in the absence of any claim regarding NIM, violation of the LGI can be used to differentiate between an oscillator described by a classical variable and a quantum mechanical oscillator \cite{15Asadian}. The classical variable does not allow the observation of SIT. Thus both the observation of SIT or LGI violations can be used to confirm the presence of states showing quantum features. In previous work, revivals and oscillations of qubit excitation in single-time detections performed over a range of settings were taken as an indirect measure for the creation of quantum superposition states \cite{96Monroe, 15Lo, 16Johnson}. These single-time detection features can be produced by the coupling to an adequate classical field distribution (see example in SI), which is not the case for LGI or SIT based on sequential measurements. 

To measure $L$ for our oscillator, we measure two-time correlations between each pair of three modular measurements $A, B, C$ while leaving out the third \cite{15Asadian}. The modular displacement settings used for the measurements can be parametrized as $\alpha_A = |\alpha|e^{i\theta_A}$, $\alpha_B = |\alpha|e^{i\theta_B}$, $\alpha_C = |\alpha|e^{i\theta_C}$, with the respective angles  $\theta_A = \omega t_1$, $\theta_B = \omega t_2$, $\theta_C = \omega t_3$ arranged to meet the constraints of successive measurements at times $t_1, t_2, t_3$. For a fixed $\alpha$ and an initial thermal state of the oscillator we numerically find values for $\theta_A, \theta_B, \theta_C$ and the phases $\phi_A, \phi_B$ and $\phi_C$ which maximize the expected value of $L$, and use these for the experiments. $L$ was measured for three thermal input states with average occupations $\overline{n}\approx 0$, $\overline{n}\approx 0.23$ and $\overline{n}\approx 0.42$. This allows testing the robustness of the protocol with respect to finite thermal occupations. Results are shown in figure \ref{fig:Correlation} (b), showing $L>1$ for displacements up to $\alpha=3$. We notice that $L$ is sensitive to noise in the experimental implementation, because it involves measuring three extremal correlations. The dashed lines in figure \ref{fig:Correlation} (b) show the expected violations for an ideal experiment and the solid lines show simulations using the level of motional and qubit dephasing which was observed in previous experiments performed in the same apparatus \cite{16Kienzler2}. Spin decoherence limits the violation at small $\alpha$, and the sharp drop in violation above $\alpha=2$ is caused by motional dephasing.

For any experiment in which SIT is observed, protocols can be designed for which $L>1$ (see SI). NSIT was previously discussed as a means to experimentally test the NIM condition \cite{13Kofler} and efforts have been undertaken to improve the LGI test by adding additional NSIT constraints and exploring their implications \cite{17Halliwell}. An alternative route is to penalize the value of $L$ by accounting for SIT between the measurements \cite{15Kujal}. The protocol we implement approaches NSIT for large displacements. If we penalize our measurement values using the theoretical value of SIT expected for our settings, only the data points highlighted by diamonds in figure \ref{fig:Correlation} (b)  produce a LGI violation (see SI for further details). For fixed displacement, the performance of the LGI protocol could be improved in the future using squeezed input oscillator states. This leads to experimentally more robust measurements, larger ratio of separation and r.m.s. wave-packet size and thus less SIT between the measurements (see SI).

The measurement techniques demonstrated here provide new tools for examining the quantum-classical divide with harmonic oscillators and could be applied in a range of experimental systems \cite{15Asadian}. Both the Leggett-Garg correlation method and SIT provide quantum signatures using few measurements, although we find experimentally that they require excellent frequency stability of the oscillator mode.  Extensions to multiple oscillators would allow tests of local realism and non-contextuality with continuous variables \cite{16Ketterer}. Alongside these fundamental applications, the combination of squeezed states and modular variable measurements demonstrated here could be used to prepare approximate GKP error-correction code states \cite{16Weigand,02Travaglione}. Ideal code states would exhibit strong SIT giving $S=1$. The control demonstrated here provides a toolbox for investigating these fault-tolerant schemes \cite{01Perskill}, opening up a new path to large-scale quantum computing with continuous variables.

\begin{acknowledgments}
We thank Thanh Long Nguyen for careful checking of the manuscript, and Peter Rabl, Renato Renner, Joseba Alonso and Maciej Malinowski for useful discussions.
We acknowledge support from the Swiss National Science Foundation under grant no. 200021 134776, ETH Research Grant under grant no. ETH-18 12-2, and from the Swiss National Science Foundation through the National Centre of Competence in Research for Quantum Science and Technology (QSIT). The research is partly based upon work supported by the Office of the Director of National Intelligence (ODNI), Intelligence Advanced Research Projects Activity (IARPA), via the U.S. Army Research Office grant W911NF-16-1-0070. The views and conclusions contained herein are those of the authors and should not be interpreted as necessarily representing the official policies or endorsements, either expressed or implied, of the ODNI, IARPA, or the U.S. Government. The U.S. Government is authorized to reproduce and distribute reprints for Governmental purposes notwithstanding any copyright annotation thereon. Any opinions, findings, and conclusions or recommendations expressed in this material are those of the author(s) and do not necessarily reflect the view of the U.S. Army Research Office.
\end{acknowledgments}

\textbf{Author Contributions:} Experimental data were taken and analyzed by CF, using an apparatus with significant contributions from VN, MM, CF. The paper was written by CF and JPH, with input from all authors. Experiments were conceived by CF and JPH.
The authors declare that they have no competing financial interests.

\iftoggle{arXiv}{
}{\bibliography{./myrefs2}}


\iftoggle{arXiv}{

\iftoggle{arXiv}{
	\clearpage
	{\LARGE\bfseries Supplementary Information\\}
}{}

We choose definitions of dimensionless position and momentum such that we have a simple connection to phase space: $\hat{X}=\sqrt{\frac{m\omega}{2\hbar}}\hat{x}$ and $\hat{P}=\sqrt{\frac{1}{2 m\omega \hbar}}\hat{p}$ leads to $\braS{\alpha}\hat{X}\ketS{\alpha}=\text{Re}(\alpha)$, $\braS{\alpha}\hat{P}\ketS{\alpha}=\text{Im}(\alpha)$ and $[\hat{X},\hat{P}]=i/2$. This definition simplifies working with position, momentum and displacement operators simultaneously.

\section{Pulse sequence realizing the asymmetric implementation}
We implement the asymmetric modular measurements making use of three internal energy levels. Besides the levels $\ket{\downarrow},\ket{\uparrow}$
we additionally use a second level in the $D_{5/2}$ manifold $\ket{a} \equiv \ket{D_{5/2}, m_j=-1/2}$. The measurement is implemented with the sequence of operations (read right to left) $\hat{R}_{1}(\phi)\hat{R}_{2}(0)\Dis{\alpha(t)\hat{\sigma}_{x,2}}\hat{R}_{2}(\pi)\hat{R}_{1}(0)$, using the definitions $\Dis{\alpha}=e^{\alpha\hat a^{\dagger} -\alpha^*\hat a}$ and $\hat{R}_{k}(\varphi) = 1/\sqrt{2} \left(\mathbb{1} - i\sin(\varphi) \hat{\sigma}_{k,x} +i \cos(\varphi) \hat{\sigma}_{k,y}\right)$. The Pauli matrices $\hat{\sigma}_{k}$ are taken to act on the $\ket{a}, \ket{\downarrow}$ basis for $k = 1$ and $\ket{\downarrow}, \ket{\uparrow}$ basis for $k = 2$. Spin rotations are implemented using resonant pulses on the two transitions, while the state-dependent displacement uses a bi-chromatic laser field resonant with both the red and blue sideband of transition $k=2$ \cite{04Haljan}.\\
If the pulse sequence is applied to an ion initially in the $\ket{\downarrow}$ level then the first pulse puts half the population in the $\ket{a}$ state. This part of the population is then not affected by the following block of operations $\hat{R}_{2}(0)\Dis{\alpha(t)\hat{\sigma}_{x,2}}\hat{R}_{2}(\pi)$ which acts on transition 2. In this block the two rotations around the SDF pulse effectively rotate the state-dependence form $\sigma_{x,2}$ to $\sigma_{z,2}$. This block therefore displaces the motion entangled with the population in $\ket{\downarrow}$. The final $\hat{R}_{1}(\phi)$ pulse then creates the state $-e^{-i\phi}\ket{\downarrow}\ket{\psi_{(+, \phi)}} + \ket{a}\ket{\psi_{(-, \phi)}}$ with $\ket{\psi_{(\pm, \phi)}}=(\mathbb{1}\pm e^{i\phi_1}\hat{\mathcal{D}} (\alpha))\ket{\psi_\text{in}}$.
We note that in the asymmetric implementation the effective qubit is given by transition 1: $\ket{\downarrow}$, $\ket{a}$.
The computational basis prior to fluorescence detection is swapped in this implementation by changing the last pulse phase $R_1(\phi)$ to $R_1(\phi+\pi)$ instead of adding an additional $\pi$-pulse.

\section{Time scales of experimental sequence}
Cooling of the calcium ion is done by precooling ($1000$ $\mu s$), Doppler cooling ($5
00$ $\mu s$), Electromagnetically Induced Transparency cooling ($400$ $\mu s$) and finally resolved sideband cooling on the axial motional mode ($250$ $\mu s$) to a mean occupation of about $\overline n \approx 0.05$ quanta. $\pi /2$-pulses on transition 2 take roughly $1.5$ $\mu s$ while on transition 1 we need around $4$ $\mu s$. Displacement operation take between $20-90$ $\mu s$. Fluorescence detection takes $60$ $\mu s$. The decision of the FPGA whether to continue with the experiment or to restart the sequence takes $50$ $\mu s$.

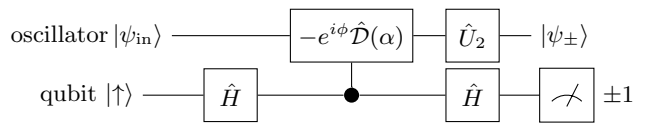
\begin{figure}[t]
\begin{tikzpicture}
\matrix[
	matrix of math nodes,
	circuit,
	column sep=1.3em,
	row sep=0.5ex,
] (m) {
\text{oscillator} \ket{\psi_{\text{in}}} & & \gate{-e^{i \phi}\hat{\mathcal{D}}(\alpha)} & \gate{\hat{U}_2}&\ket{\psi_\pm}\,\\
\text{qubit $\ket{\uparrow}$}& \gate{\hat{H}} & \C & \gate{\hat{H}} & \meas\\
};
\begin{scope}
\draw (m-1-1) -- (m-1-3)-- (m-1-4)--(m-1-5);
\draw (m-2-1) -- (m-2-2) -- (m-2-3) -- (m-2-3) -- (m-2-4)-- (m-2-5);

\begin{scope}[every path/.style={double}]
\end{scope}

\begin{scope}[every path/.style={double,double distance=0.5ex}]
\end{scope}

\draw (m-1-3) -- (m-2-3);

\end{scope}

\node[right] at (m-2-5.east) {$\pm 1$};
\end{tikzpicture}
\caption{Circuit model of the implemented modular position and momentum measurements. $\hat{H}$ denotes the Hadamard gate, $\hat{\mathcal{D}}(\alpha)$ the displacement operator, and the qubit is measured projectively in the $\hat{\sigma}_z$ basis. 
We experimentally investigate the asymmetric implementation $\hat{U}_2 =\mathbb{1}$,
the symmetric implementation $\hat{U}_2 =\hat{\mathcal{D}} (-\alpha/2)$ and free oscillator evolution $\hat{U}_2(t) =e^{-i\omega t a^\dagger a}$. 
}
\label{circuits}
\end{figure}

\section{Calibration of modular measurements}
We calibrate the SDF pulse and perform two additional laser phase calibrations. These together with automated calculations of phases due to Stark-shifts using the known pulse durations and timings allows us to run in-principle arbitrarily long sequences of modular measurements.

\subsection{SDF pulse}
The SDF pulse is calibrated by first roughly balancing blue- and red-sideband powers and applying it to an initial ground state cooled oscillator in the $\ket{\downarrow}$ internal level for a time $t_\text{SDF}$. The decrease of $P(\downarrow)$ probability is observed and iteratively we find better balanced laser powers and a Stark-shift of our transition by smoothing out the $P(\downarrow)$ signal at 0.5 probability for the timescales required in the experiment. From the calibrated SDF pulse we extract the proportionality factor between pulse time and displacement size by fitting the qubit decay to its expected form $P(\downarrow)=\frac{1}{2}(1+e^{-2(c t_{\text{SDF}})^2})$ with $c$ the floated proportionality constant. Typical values of $c$ obtained from these fits are: $c \approx 0.035-0.028$ ${\mu s}^{-1}$  An example of this calibration is given in figure \ref{fig:Calibrations} (a).

\subsection{laser-transition phase evolution}
To realize the modular measurements in the asymmetric implementation we need the relative laser-transition phases as well as the oscillator time evolution to be phase-locked. The SDF pulse addresses transition 2 with a bi-chromatic pulse where the average of this two frequencies addresses transition 2 while the difference acts on the motional space:  $\hat H_{SDF}=\frac{\eta \hbar \Omega}{2}(\hat \sigma_+e^{-i\overline{\phi}} + \hat \sigma_-e^{i\overline{\phi}})(\hat a e^{i\Delta\phi/2+i\delta t}+\hat a^{\dagger}e^{-i\Delta\phi/2-i\delta t})$ with the phases $\overline{\phi}$ and $\Delta\phi$ given by the average and difference of the blue and red sideband laser phases. The average frequency when addressing transition 2 with the SDF differs by the calibrated Stark-shift to the frequency used in a resonant carrier pulse. To account for this mismatch in phase evolution we match the phase of the first $R_2$ rotation to the SDF and calculate the phases of following $R_2(\varphi)$ pulses based on the calibrated Stark-shift and the timing of our sequence. The action of the calibrated block $\hat{R}_{2}(0)\Dis{\alpha(t)\hat{\sigma}_{x,2}}\hat{R}_{2}(\pi)$ on $\ket{\downarrow}$ gives a constant probability $P(\downarrow)\equiv 1$ independent of the displacement size. An example using this characteristic for calibration is given in figure \ref{fig:Calibrations} (b).

\subsection{Superposition phase controlled by $R_1(\phi)$ pulse}
The phases of transition 1 pulses $R_1(\phi)$ are solely calibrated on the expected physics. We note that after each modular measurement the qubit is detected and projected to $\ket{\uparrow}$. Therefore in contrast to the oscillator the laser-transition 1 and 2 phase is reset at the start of each modular measurement. Thus for two sequential measurements with the same duration of the SDF pulse the phase is identical. A single modular measurement with a general displacement $\alpha$ does not allow to calibrate $\phi$. This is because for large enough displacements $P(\downarrow)\equiv 0.5$ which is independent of $\phi$. Instead we calibrate the phase $\phi$ by a correlation measurement with settings $\alpha_A = -\alpha_B$, varying $\phi_A$ and $\phi_B$ jointly and fitting to the theoretical expectation. An example is given in figure \ref{fig:Calibrations} (c).

\section{Comparison of the two implementations}
The symmetric implementation besides its simpler pulse sequence has several further advantages: The transition $\ket{\downarrow}, \ket{\uparrow}$ has $ \approx 6$ times more laser power available and half the magnetic field sensitivity compared to qubit used in the asymmetric implementation $\ket{\downarrow}, \ket{a}$. Therefore whenever possible we use the symmetric implementation. Both implementations can be represented in the circuit model by the circuit given in figure \ref{circuits}.

\section{Qubit readout}
In a temporal sequence of measurements the last measurement is performed as a long fluorescence detection with a typical readout time of 200 $\mu s$. All the preceding measurements are performed with a shorter fluorescence time of 60 $\mu s$ having a detection error of $\epsilon_{\text{short}}\approx 4\cdot 10^{-3}$ and giving an average of roughly 10 counts for a bright detection result. Our imaging system collects $4.4 \%$ of the emitted photons and the PMT quantum efficiency is roughly $26.5\%$. This amounts in an average scattering of 1000 photons from the ion for this shorter detection.

\section{NSIT, Observables, Kraus operators and Commutators}
If the order of measurements $A, B$ does not matter then for projective measurements their respective observables commute. For the generalized measurements considered in this work this no longer holds which we can see by the following argument. Still there exists a general relation between Kraus operators and NSIT: $\langle [\hat{E}_{a}^A, \hat{E}^{\dagger B}_{b} \hat{E}_{b}^{B}]\rangle= 0 \Rightarrow \text{NSIT}$. This relation is derived here:
\begin{align}
P_{B(A)}(b) 
&= \sum_{a} P_{BA}(b,a)\\ \notag
&= \sum_{a} {\rm Tr} \left\{ \hat{E}^{\dagger A}_{a} \hat{E}^{\dagger B}_{b} \hat{E}^{B}_{b} \hat{E}^{A}_{a} \hat{\rho}_i \right\}
\end{align}
Using $\langle[\hat{E}_{a}^A, \hat{E}^{\dagger B}_{b} \hat{E}_{b}^{B} ] \rangle = 0$ this translates into
\begin{align}
P_{B(A)}(b) &= \sum_{a} {\rm Tr} \left\{ \hat{E}^{\dagger A}_{a} \hat{E}^{A}_{a} \hat{E}^{\dagger B}_{b} \hat{E}^{B}_{b} \hat{\rho}_i \right\}\\\notag
&= {\rm Tr} \left\{\sum_{a} \hat{E}^{\dagger A}_{a} \hat{E}^{A}_{a} \hat{E}^{\dagger B}_{b} \hat{E}^{B}_{b} \hat{\rho}_i \right\}\\ \notag
&= {\rm Tr} \left\{ \hat{E}^{\dagger B}_{b} \hat{E}^{B}_{b} \hat{\rho}_i \right\}\\ \notag
&=P_B(b)
\end{align}
Straight forward calculation of the commutator $[\hat{E}_{a}^A, \hat{E}^{\dagger B}_{b} \hat{E}_{b}^{B}]$ for the two implementations as well as the commutator of the modular observables $[\hat{Q}_A,\hat{Q}_B]$ leads to the conditions for NSIT or commutation of the observables given by:
\begin{align}
&\text{Sym.}: &\text{Im}(\alpha_B\alpha_A^\ast)&=2\pi k_1 , &k_1\in\mathbb{Z}&\Rightarrow \text{NSIT}\\\notag
&\text{Asym}:&\text{Im}(\alpha_B\alpha_A^\ast)&=\pi k_2 , &k_2\in\mathbb{Z}&\Rightarrow \text{NSIT}\\\notag
&[\hat{Q}_A,\hat{Q}_B]:&\text{Im}(\alpha_B\alpha_A^\ast)&=\pi k_3 , &k_3\in\mathbb{Z}&\Rightarrow [\hat{Q}_A,\hat{Q}_B]=0\notag
\end{align}
Therefore if the commutator of the observables vanishes with an odd $k_3$ number then $\text{Im}(\alpha_B \alpha_A^\ast)=\pi k_{\text{odd}} \neq 2 \pi k_1$, thus the symmetric implementation is SIT. This is a general case for which the observables commute but the sequential measurements are SIT. In figure \ref{fig:AdditonalSIT} additional data for SIT and NSIT experimental sequences are shown. In particular one can compare the two different implementations and see that NSIT does not imply commutation of the observables.

\section{Theoretical Wigner function plots of experimentally created states}
In the main text figure \ref{fig:NSIT} (c) we plotted the Wigner functions of the states created during our measurement of NSIT for modular position and momentum for one example of input state. The chosen state was a superposition of a ground state cooled oscillator but the same experiment was also performed with either an oscillator in a squeezed state or in the first excited state. In figure \ref{fig:addStates} the equivalent plot as in the main text with the same orientation of initial superposition but now based on a (a) first excited or (b) a squeezed state are shown. 

\section{Correlator full data set}
The full measurement data of the correlation measurements presented in figure \ref{fig:Correlation} (a) of the main text are shown in figure \ref{fig:CorrelationThird}. Besides the measured correlators also the probabilities measured in the first measurement $A$ and the second measurement $B$ are shown. 

\section{Leggett-Garg violation measurement settings}
In order to find the measurement settings with which we violate the LGI we calculate the analytic expression for the value of $L$ depending on the initial ion temperature and the displacement size $\alpha$ as well as the measurement settings A: $(\theta_1, \phi_A)$, B: $(\theta_2,\phi_B)$ C: $(\theta_3,\phi_C)$. For each temperature and displacement $\alpha$ we maximize the found analytic expression over $\theta_1,\theta_2, \theta_3$ and $\phi_A, \phi_B, \phi_C$ using Mathematica. To do so we first find a local maximum for a small displacement $\alpha=0.2$, then we use the settings found from this analysis as an initial guess for the maximization for a slightly larger displacement $\alpha=0.25$ like this we find successively the settings for larger displacements. In figure \ref{fig:RawDataLGViolation} some raw data of $L$ violations measurements together with the used experimental settings are shown.

The temperature of the oscillator is calibrated by shortening the cooling sequence used and subsequently reading out the Fock state populations of the oscillator fitting them to a thermal state of the oscillator. Before each Leggett-Garg experiment the phase $\phi$ is calibrated for the displacement size and temperature in the manner described before. Based on this single calibration the three correlations are measured.

\section{Effect of noise on Leggett-Garg inequality violations}
The motional dephasing is accounted for by solving the Lindblad master equation during the state-dependent-force pulses with a dephasing operator $\sqrt{30}(\hat{a}\hat{a}^{\dagger}+\hat{a}^{\dagger}\hat{a})$ with 30 dephasing jumps/s. The line-with of  the transition 1 is known from Ramsey measurements to be $l\approx 665 \text{ Hz FWHM}$ and varies on timescales longer than an experimental shot thus we include it by averaging over 4000 randomly chosen phases $\phi$ from a normal distribution with $\sigma = \frac{l \pi t_{SDF}}{\sqrt{2 \ln(2)}}+0.087$ where the last term accounts for phase calibration errors. 

\section{Penalized LGI}
The paper by Kujala et al. \cite{15Kujal} considers cyclic contextuality inequalities. Some LGIs, for example $\widetilde{C}_{AB}+\widetilde{C}_{BC}+\widetilde{C}_{CA}<1$, are special cases of these cyclic contextuality inequalities. The paper addresses the problem that even if an experimenter intends to perform compatible measurements due to experimental fluctuations and imprecision’s, there will still be a certain amount of SIT between the sequential measurements performed. The work derives penalized contextuality bounds to account for these imprecision’s. The derived penalization is expressed in the notation of this work $TS=2(|\widetilde{S}_{AB}|+|\widetilde{S}_{BC}|+|\widetilde{S}_{CA}|)$ and can be interpreted as a total amount of SIT observed. They consider cyclic measurements, thus each measurement is performed once as the first measurement in the sequence and once as a second measurement. This allows the penalization to be extracted directly from the contextuality bound measurements.

In contrast, the inequality we considered in this work is not cyclic $C_{AB}+C_{BC}-C_{AC}<1$. Only $B$ is once performed as a first measurement and another time as a second measurement. From this we extract $\widetilde{S}_{AB}$. Performing an additional measurement of $C$ directly on the input state would allow to extract $|\widetilde{S}_{BC}|$ and $|\widetilde{S}_{AC}|$ and then to calculated a penalized $L$ value: $L_{\text{pen.}}=L-2(|\widetilde{S}_{AB}|+|\widetilde{S}_{BC}|+|\widetilde{S}_{AC}|)=L-TS$.

In this penalization we assumed that there is no backward SIT. This means that if we perform first a measurement $A$ then $B$ that $P_{A}(a)=P_{(B)A}(a) \equiv \sum_b P_{BA}(b, a)$, which in a real experiment again will only be approximately given. The penalization for the cyclic inequalities also contain this type of fluctuation to some extent. Further, there might be subtleties which we miss at this stage.

Our LGI protocol is based on performing incompatible measurements thus we expect SIT between the measurements. Analytic calculation of the amount of inbuilt SIT in our protocol, see figure \ref{fig:Penalisation} (a), shows that this is indeed the case for displacements of around $\alpha=1$. But the amount of SIT approaches zero for the larger displacements. From this we conclude that for large displacements the ideal protocol approaches $L$=1.5 with NSIT measurements at the two-time level.
If we subtract the theoretical amount of inbuilt SIT from our data, see figure \ref{fig:Penalisation} (b), then at our experimentally achieved size of displacements some points violate the LGI in this penalized fashion. These are the points highlighted with diamonds in the main part of the paper. Further, we can get a feeling for how close our experiment resembles the theoretical amount of inbuilt SIT, by extracting $|\widetilde{S}_{AB}|$ from our experimental data, see figure \ref{fig:Penalisation} (c). The theoretical expectation for $|\widetilde{S}_{AB}|$ is around zero and never exceeds 0.02. The amount of SIT we measure is close to zero but slightly higher than this theoretical expectation.  The higher amount of SIT is expected given the accuracy with which we can calibrate and perform our experiments.
A number of methods have been proposed for performing a LGI test using NSIT measurements, see for example \cite{13Kofler} where the key is to use mixed input states. In the work \cite{17Halliwell} a LGI test using two time NSIT measurements is called a test of an intermediate form of MR.

\section{LGI and SIT as efficient quantum witnesses}
This work shows the violation of an LGI using a mechanical oscillator, a system which allows to explore the quantum to classical transition in a natural way.
Further, we explore SIT as an alternative quantum witness. Both require few measurements for the confirmation of the quantum states: LGI needs 12 fluorescence detections while SIT needs 6. This is much less than we typically require to extract a negative Wigner function point. For the latter we extract the Fock state populations of the oscillator from a sideband flopping curve which requires around 200 fluorescence detections \cite{16Kienzler}.
We find that that the LGI methods require excellent frequency stability of the mechanical oscillator under test. The SIT quantum witness has the advantage of involving only measurements at two times. But SIT has the need of $m_{\alpha_B}$ to be non zero, which requires more involved oscillator input states.

\section{Ramsey analogy and advantage of sequential measurements}
The modular measurements presented can be viewed as Ramsey measurements coupling to a quantum field which is given by the oscillator phase space. Thus the best semi-classical comparison is given by a Ramsey measurement coupling to a classical variable $x(t)$: $\hat{H}\propto \ketS{\uparrow}\braS{\uparrow} x(t) \propto \ketS{\uparrow}\braS{\uparrow} (\hat{a}e^{i\Delta\phi/2} +\hat{a}^\dagger e^{-i\Delta\phi/2})$. An extensive discussion of this comparison can be found in the supplemental material of \cite{15Asadian}.

In previous experiments with superposition states, single detection results were used to confirm the creation of superposition states \cite{96Monroe,15Lo,16Johnson}. Such single time detection results could in principle emerge from the coupling to a classical variable $x(t)$. In particular if $x(t)$ contains dominant frequency components a variety of oscillations and revivals in the qubit probabilities can be observed. As illustration we consider a simple example: $x(t)=A \cos(2\pi f t)$ is given by a single frequency component $f$ with a fluctuating amplitude $A$. The amplitude fluctuates on slow time scales compared to a single experimental shot and its probability distribution is given by the Gaussian $P(A)=\frac{1}{\sigma \sqrt{2\pi}}e^{-\frac{1}{2}(\frac{A-A_0}{\sigma})^2}$, and the experiment is synchronized with respect to the noise frequency (an example of such a noise source would be noise due to the mains lines measured by a line triggered experiment). For such a periodic noise source we find $\langle\hat{Q}\rangle=\langle P(+1)-P(-1)\rangle =-e^{\frac{1}{2}(\frac{\sin(2\pi f T)\sigma}{2\pi f})^2} \cos(\phi+\frac{A_0}{2\pi f} \sin(2 \pi f T))$ where $T$ is given by the Ramsey interaction time and $\phi$ is the phase of the second Ramsey pulse. $\langle\hat{Q}\rangle$ is plotted in figure \ref{fig:ClassicalOscillations}, where we can see that it exhibits very similar oscillations to those observed in experiments like \cite{96Monroe}. Single measurements thus have a hard time proving that the experiments actually create Schr{\"o}dinger cat like superpositions. Nature could be malicious and one could always just couple to a classical variable $x(t)$ giving rise to the observed oscillations and revivals.

The distinction between the coupling to a classical field or a quantum field is easier when considering sequential measurements. In the quantum case the first measurement creates a superposition state of the quantum field which can change the statistics of the second measurement (SIT). In the classical case the variable $x(t)$ is not changed by the first measurement and SIT will not be observed between two measurements. 

\section{SIT measurements violating a Leggett-Garg inequality}
Here we give an explicit procedure how to violate the Leggett-Garg inequality $L= C_{AB} + C_{BC} - C_{AC}\leq 1$ having observed SIT between two modular measurements on an input state. This procedure is equivalent to the one used and discussed in \cite{15Robens} and is briefly commented on in \cite{14Emary}.\\
We consider two measurements $B$, $C$ which we read out by coupling them to an ancilla qubit. Thus the measurement of the qubit has two possible outcomes, Up and Down, which we label U and D. The only assumption we make about the measurements $B$ and $C$ is that $B$ is SIT to $C$ when the input state $\ket{\psi}$ is measured. This means  $P_C(c)\neq \sum_{b} P_{CB}(c,b) = P_{C(B)}(c)$. Thus one of the two probabilities needs to be bigger than the other. Without loss of generality we choose $P_C(D)< P_{C(B)}(\text{D})$ and we define $a \equiv P_{C(B)}(\text{D})-P_C(\text{D})$ to be the difference between the two. (In the case of $P_C(\text{U})> P_{C(B)}(\text{U})$ we can modify the protocol slightly). Measurement $A$ is simply the state preparation or confirmation of the state preparation of $\ket{\psi}$. The key point of the protocol is to assign different measurement results to the outcomes U, D in each of the measurements $A,B,C$. The assigned results are always $r=\pm 1$  which is compatible with the assumption $|r|\leq1$ used in the  proof of the LGI \cite{14Emary}. To be specific the three measurements violating the LGI are given by:\\
\begin{description}
	\item[$t_0$] Initial state preparation of $\ket{\uparrow}\ket{\psi}$
	\item[$t_1$] Measurement $A$: Readout of the qubit. We assign the result U the value +1 and -1 to D; $f_A(\text{U})=+1, f_A(\text{D})=-1$
	\item[$t_2$] Measurement $B$ and both results D and U are identified with +1; $f_B(\text{D})=f_B(\text{U})= +1$
	\item[$t_3$] Measurement $C$. Result U: -1 Result D: +1; $f_C(\text{U})=-1, f_C(\text{D})=1$
\end{description}
Note that the assignment of a constant value for measurement $B$ can be interpreted as performing the measurement but not looking at the result. We can now calculate the violation of the LGI. $C_{AB}=1$ since measurement $A$ always gives an Up result and in measurement $B$ we assigned the constant value of +1.
The correlator $C_{AC}=\sum_{c} f_C(c) P_{C}(c)$ simplifies to be the expectation value of measurement $C$ since measurement $A$ is only confirming the state preparation. $C_{BC}=\sum_{c} f_C(c) \sum_{b} P_{CB}(c,b)$ simplifies since we assigned in measurement $B$ the constant value of +1
Thus:
\begin{align}
L&=1+(\sum_{c} f_C(c) \sum_{b} P_{CB}(c,b) - \sum_{c} f_C(c) P_{C}(c))\\\notag
&=1+\sum_{c} f_C(c) (\sum_{b} P_{CB}(c,b) - P_{C}(c))\\\notag
&=1+2a>1.
\end{align}
We can also see that in the case of $P_C(\text{D})> P_{C(B)}(\text{D})$ we can change the assignment of results in measurement $C$ to $f_C(\text{U})=1, f_C(\text{D})=-1$.

\section{GKP state allowing S=1}
This can be seen in various ways lets consider the formula for $S$ in the symmetric implementation:
$S = \frac{1}{2}\left(1-\cos(\Phi)\right) |m_{\alpha_B}| \cos(\arg(m_{\alpha_B}))$.  $S$=1 requires $\Phi=\pi$ and $m_{\alpha_B}=1$. This is fulfilled if we choose a GKP input state with periodicity $\alpha_B \in \mathbb{R}$: $\sum_{l=- \infty}^{\infty} \Dis{l \alpha_B}\ket{x=0}$ and $\alpha_A= i \pi /{\alpha_B}$.

\section{Potential improvement of LGI violation using squeezed oscillator input states}
We maximize the $L$ value for squeezed initial oscillator states $\hat{S}(\xi)\ket{0}$ with $\hat{S}(\xi)=e^{(\xi^*\hat{a}^2-\xi\hat{a}^{\dagger^2})/2}, \xi=r e^{i\phi}$ the squeezing operator and parameter. Comparing a squeezed state with $r\approx 0.9$ to a ground state we find that with the same displacement size the squeezed state allows higher violation see figure \ref{fig:SqLG} (a). The simulation of realistic dephasing noise, figure \ref{fig:SqLG} (b) shows that this advantage is still present in a realistic scenario. Further the SIT inbuilt in the measurements drops much quicker for the squeezed state. Analytic calculations of the inbuilt SIT are shown in figure \ref{fig:SqLG} (c). Also the created states for a fixed $|\alpha|$ are in some sense more macroscopic since the ratio of separation to relevant wave-packet extend (approximately the squeezed axis for larger $\alpha$) is larger. For $r=0.9$ the ratio is improved by a factor of $\approx 2.5$.

\section{Data availability}
The data that support the plots within this paper and other findings of this study are available from the corresponding authors on request. 

\makeatletter
\apptocmd{\thebibliography}{\global\c@NAT@ctr 28\relax}{}{}
\makeatother

\iftoggle{arXiv}{
}{\bibliography{./myrefs2}}

\newpage

\begin{figure*}[htbp]
	\resizebox{1\textwidth}{!}{\includegraphics{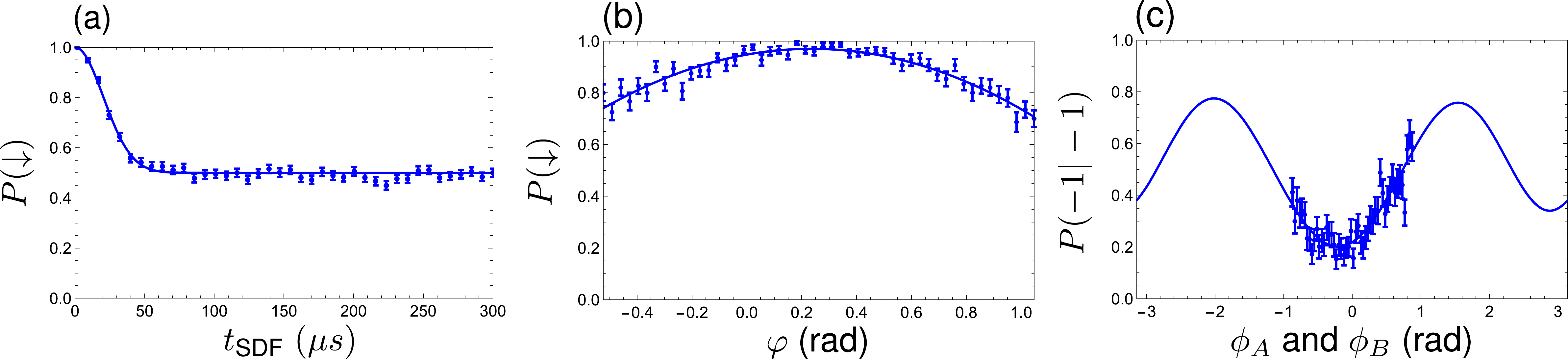}}
	\caption{Calibrations: (a) shows a calibrated SDF pulse applied to a ground state cooled ion and a fit to the expected behavior $P(\downarrow)=\frac{1}{2}(1+e^{-2(c t_{\text{SDF}})^2})$. From this fit we extract the proportionality constant $c$ between displacement size and $t_{\text{SDF }}$ which we use in all analytic calculations of expected measurement results. (b) shows the phase matching between addressing transition 2 with single frequency or with a bi-chromatic pulse. The sequence consist out of a single $\pi / 2$-pulse followed by an SDF pulse here with $t_{\text{SDF }}=200\:\mu s$. Whenever the $\pi /2$-pulse phase $\varphi$ is matched to the SDF no superposition will be created instead the full wave-packet gets displaced and therefore the $P(\downarrow)\equiv1$ for any SDF duration. Finally we need to calibrate the phase of the $R_1(\phi)$ pulse using an experimental calibration shown in (c). We use a sequence of two modular measurements with the same duration and opposite SDF directions in order to be able to observe a signal. We sweep both phases $\phi_A$ and $\phi_B$ simultaneously and readout the relevant minimum.}
	\label{fig:Calibrations}
\end{figure*}

\begin{figure*}
	\resizebox{1\textwidth}{!}{\includegraphics{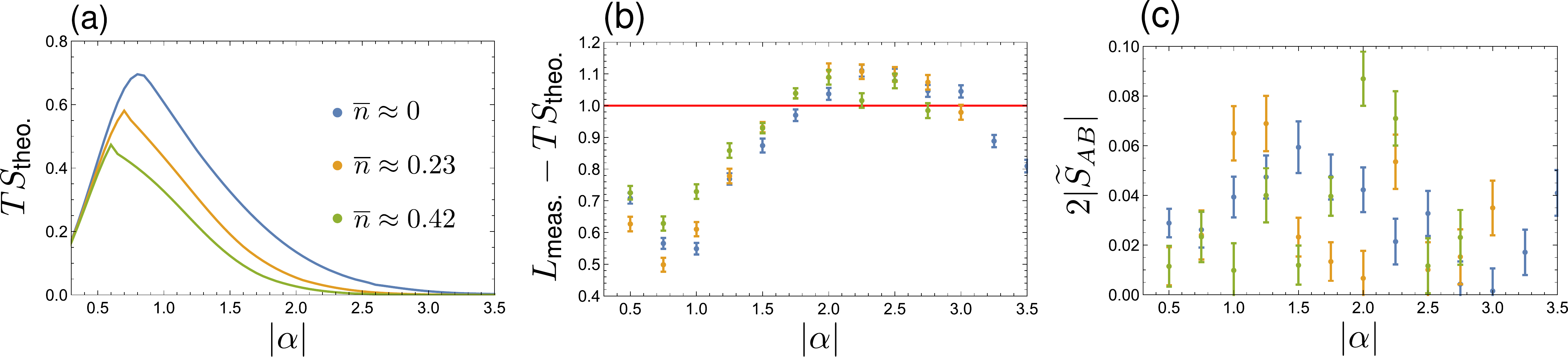}}
	\caption{Penalization of $L$ for inbuilt two-time SIT. (a) Analytic calculation of the total amount of SIT $ST=2(|\widetilde{S}_{AB}|+|\widetilde{S}_{BC}|+|\widetilde{S}_{AC}|)$ due to incompatible settings used to violate the LGI. $ST$ approaches zero for the larger displacements $\alpha$. (b) The measured data penalized by the theoretical amount of inbuilt SIT. Several points around $\alpha$ 2.25 are still able to violate the LGI in this penalized fashion. (c) $\widetilde{S}_{AB}$ extracted from the experimental data. The amount of SIT is higher than expected from the analytic calculation, which predicts values up to $0.02$. But the values stay close to zero. Note that the the total SIT is dominated by $|\widetilde{S}_{BC}|$ and $|\widetilde{S}_{AC}|$.}
	\label{fig:Penalisation}
\end{figure*}

\begin{figure*}
	\resizebox{1\textwidth}{!}{\includegraphics{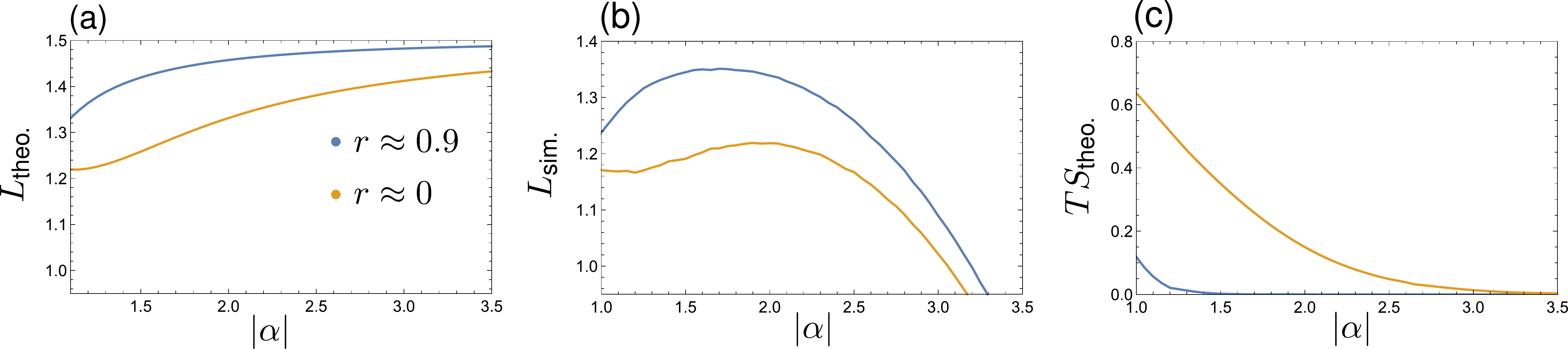}}
	\caption{Theoretical comparison of a LGI experiment using a ground state cooled or a squeezed oscillator state. Results are presented for a squeezing parameter $r\approx 0.9$. (a) Analytic calculation of the achievable $L$ values. (b) Simulated $L$ values including the same motional dephasing and line-widths as in the simulations of figure \ref{fig:Correlation} (b) in the main paper. (c) Analytic calculation of the amount of $ST$ present during the measurements.}
	\label{fig:SqLG}
\end{figure*}

\begin{figure*}
	\resizebox{1\textwidth}{!}{\includegraphics{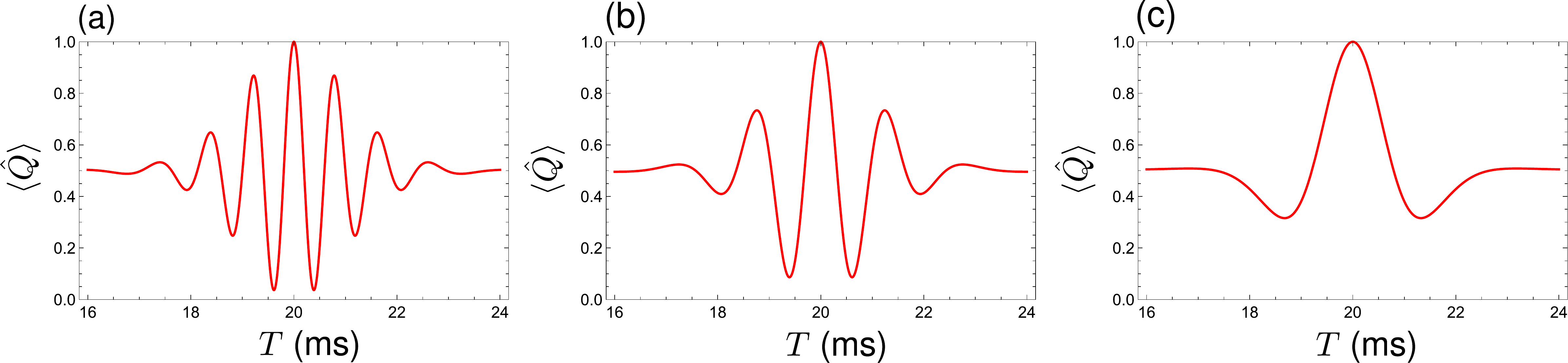}}
	\caption{Expected results for a Ramsey measurement coupling to $x(t)=A \cos(2\pi f t)$ a classical single frequency noise source.
	The amplitude $A$ of this single frequency noise fluctuates on timescales slower than an experimental shot with a Gaussian probability distribution $P(A)=\frac{1}{\sigma \sqrt{2\pi}}e^{-\frac{1}{2}(\frac{A-A_0}{\sigma})^2}$. The coupling constant is assumed to be 1: $\hat{H}= \ketS{\uparrow}\braS{\uparrow} x(t)$. Thus we find $\langle\hat{Q}\rangle=\langle P(+1)-P(-1)\rangle =-e^{\frac{1}{2}(\frac{\sin(2\pi f T)\sigma}{2\pi f})^2} \cos(\phi+\frac{A_0}{2\pi f} \sin(2 \pi f T))$ with $T$ given by the Ramsey wait time and $\phi$ the second Ramsey pulse phase. Shown are 3 different amplitudes $A_0$: (a) $A_0$ = 8000, (b)  $A_0$ = 5000 and (c) $A_0$ = 2000 of the noise with the noise frequency fixed to 50 Hz, $\sigma= 1000$, $\phi=0$. The oscillations resemble characteristic traces of Schr{\"o}dinger cat states, such as those found in \cite{96Monroe}.}
	\label{fig:ClassicalOscillations}
\end{figure*}

\begin{figure*}
	\resizebox{1\textwidth}{!}{\includegraphics{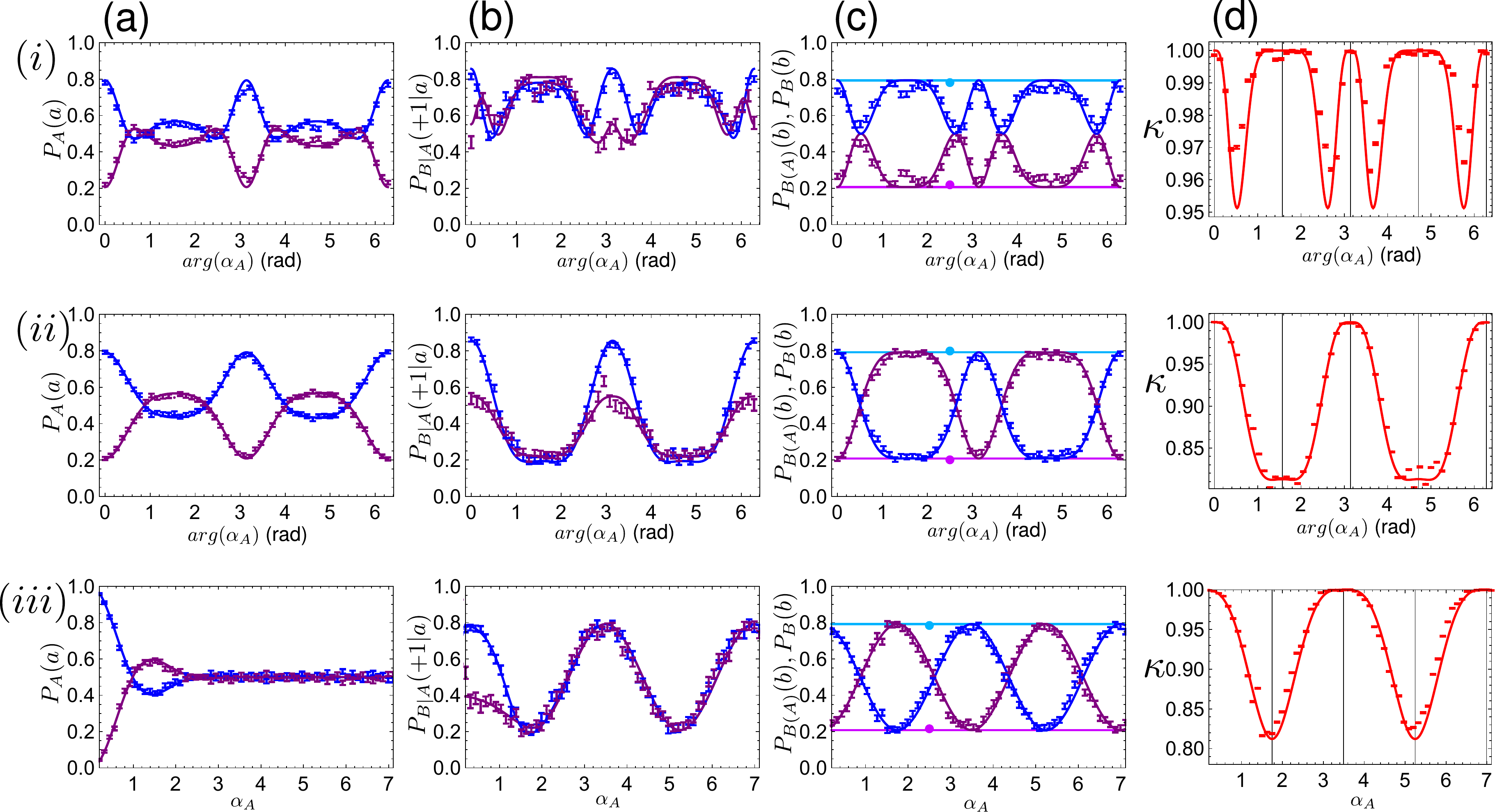}}
	\caption{Additional SIT and NSIT data compared to observable commutation. Column (a) and (b) show the direct detection data obtained in measuring first $A$ and then $B$. The blue line corresponds to $a=+1$ while purple stands for $a=-1$, column (c) show the result of measuring $B$ alone. $P_B$ is given as a single point plotted at $arg(\alpha_A)=2.5$ and it is compared to $P_{B(A)}$ which is calculated from the results of columns (a) and (b). Blue: $b=+1$ while purple: $b=-1$. Finally the classical fidelity $\kappa \equiv \sum_{b}\sqrt{P_B(b)P_{B(A)}(b)}$ \cite{06Ralph} is shown in column (d). The classical fidelity can be used as an alternative to $S$ in order to quantify the amount of SIT. For NSIT measurement $\kappa=1$. The black vertical lines in (d) indicate settings for which the underlying observables commute. $(i)$: Asymmetric implementation, SIT as function of $arg(\alpha_A)$ with $|\alpha_A|=\sqrt{\pi}$, $\alpha_I=\sqrt{\pi}, \alpha_B=-\sqrt{\pi}$, $\phi_A=\phi_B=0$. The data varies between SIT and NSIT, where NSIT is observed for the settings where the observables commute. $(ii)$: The same experiment with the symmetric implementation, $\phi_A=\phi_B=\pi$ and we observe that this implementation can be SIT even in cases where the observables commute. $(iii)$: SIT as function of $\alpha_A$ in the symmetric implementation with $\alpha_B = \alpha_I\approx\sqrt{\pi}$. Again SIT is observed at points where the observables commute.}
	\label{fig:AdditonalSIT}
\end{figure*}

\begin{figure*}
	\resizebox{1\textwidth}{!}{\includegraphics{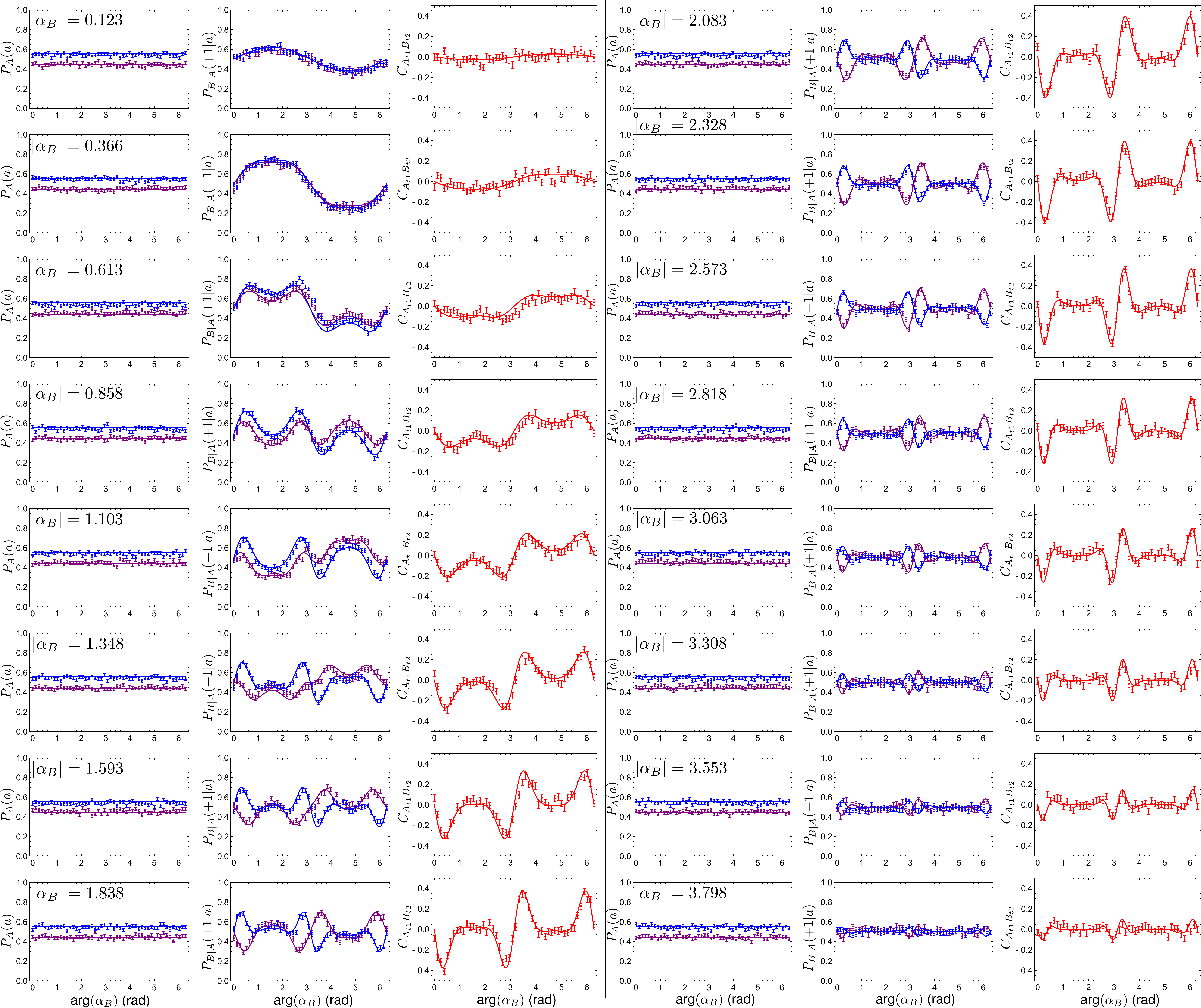}}
	\caption{Two-time correlation measurement in the asymmetric implementation. The fixed experimental settings were $\alpha_A=2.1$, $\phi_A=0$, $\phi_B=\pi/2$. Given in blue: $a=+1$ while purple shows: $a=-1$. Solid lines show the expectations for an ideal experiment. Error bars are given as standard deviations of the mean errors (SEM) $P_A(a), P_{B|A}(+1|a)$ and are propagated from theses for the $C_{AB}$ errors. The data sets shown here create together the 3D plot of figure \ref{fig:Correlation} (a) in the main text.}
	\label{fig:CorrelationThird}
\end{figure*}

\begin{figure*}
	\resizebox{1\textwidth}{!}{\includegraphics{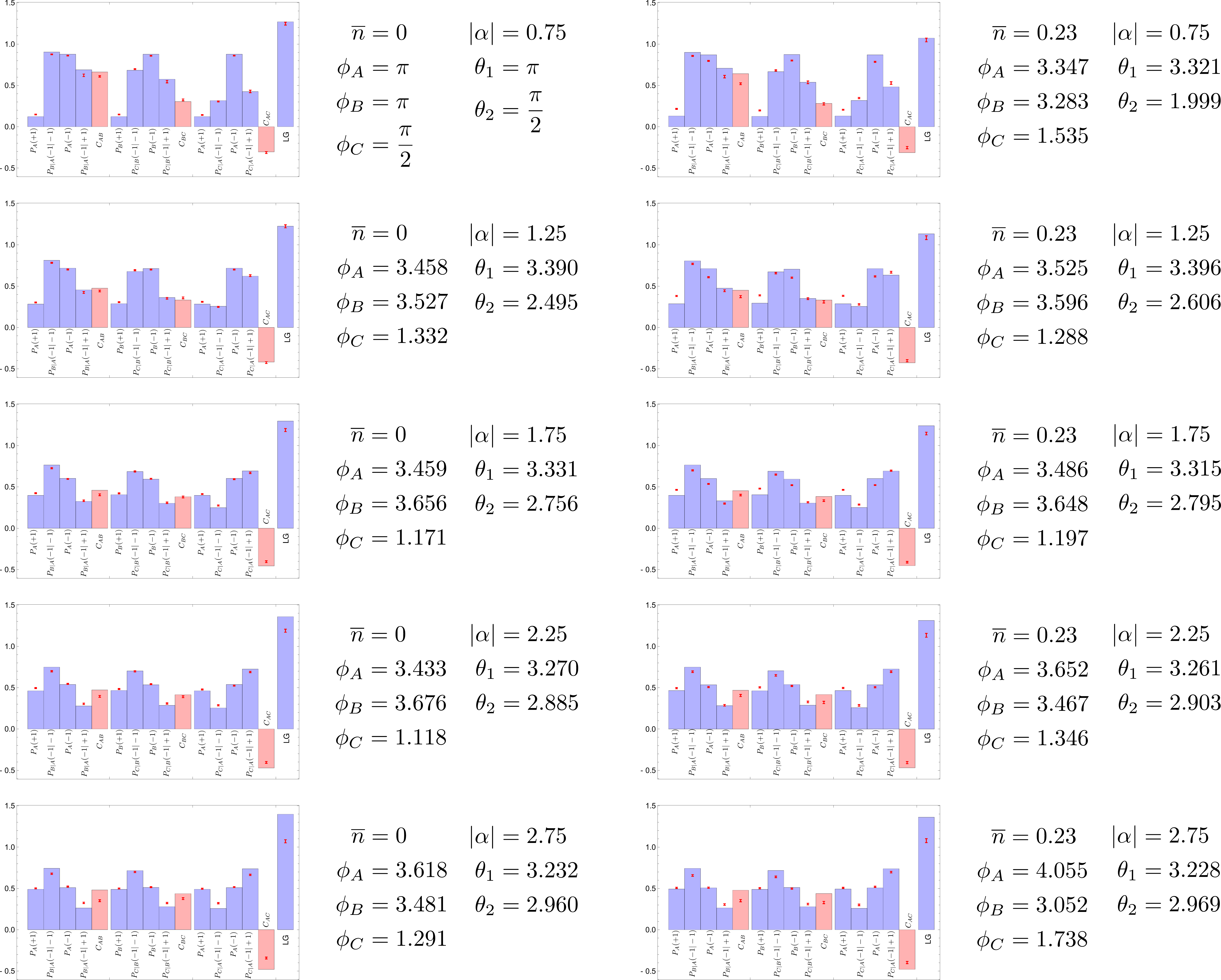}}
	\caption{A sample of individual detection data and measurement settings for the Leggett-Garg violations presented in figure \ref{fig:Correlation} (b) of the main text. The data is from the set $\overline{n}\approx0$ and $\overline{n}\approx0.23$. Data is shown as red points while the expectations for an ideal experiment are shown as bars. Blue bars are the detection data and the three red bars show the correlations calculated from these detections. We see qualitatively good agreement for smaller $\alpha$ which decreases for higher displacements, which is mainly due to dephasing noise in the experimental system. The settings $\theta_1$ and $\theta_2$ correspond to  $\theta_1 = \theta_B-\theta_A$, $\theta_2 = \theta_C-\theta_B$.}
	\label{fig:RawDataLGViolation}
\end{figure*}

\begin{figure*}
	\resizebox{0.9\textwidth}{!}{\includegraphics{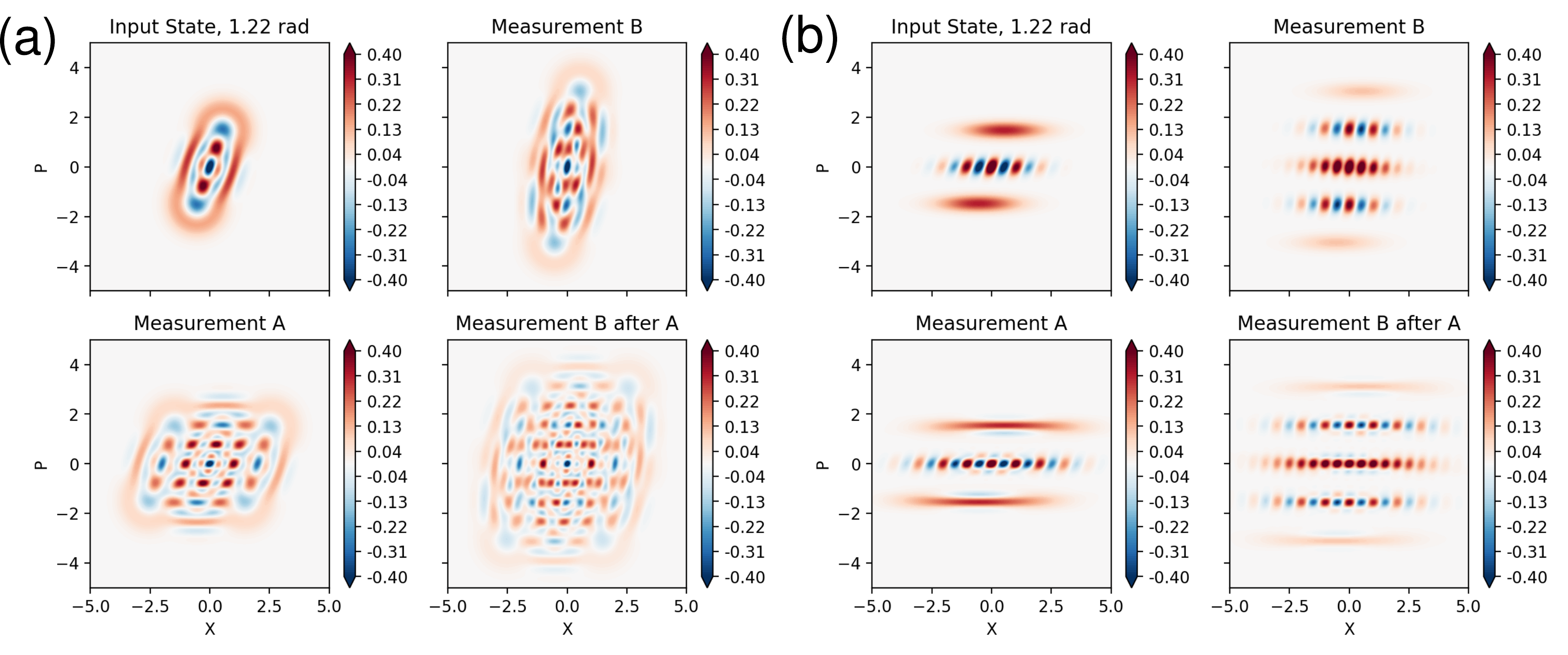}}
	\caption{Theoretical Wigner function plots of two more examples of input states for the measurements presented in figure \ref{fig:NSIT} of the main text. The displayed input states are  $(\hat{\mathcal{D}}(-|\alpha_B|e^{i\phi_I}/2) + \hat{\mathcal{D}}(|\alpha_B| e^{i \phi_I}/2))\ket{\phi}$ with $\phi_I=1.22 \text{ rad}$ and (a): $\ket{\phi}$=$\ket{1}$ and (b): $\hat{S}(-0.82)\ket{0}$. Further their post-measurement states with result $+1$ during the experimental sequence are shown.}
	\label{fig:addStates}
\end{figure*} 

}{}

\end{document}